\documentstyle[floats,aps]{revtex}

\newcommand{\forget}[1]{\iffalse#1\fi}
\newcommand{\forgetmenot}[1]{\iftrue#1\fi}

\newcommand{\be}{\begin{equation}}
\newcommand{\ee}{\end{equation}}

\renewcommand{\:}[2]{{\textstyle\frac{#1}{#2}}}

\newcommand{\bra}[1]{\left(#1\right)}

\newcommand{\udot}{\dot u}
\newcommand{\del}{\nabla}
\newcommand{\sdel}{\widetilde\del}

\renewcommand{\>}{\rangle}

\newcommand{\etal}{{\it{et~al.}}}

\title{Magnetic Fields and the Cosmic Microwave Background}
\author{C. A. Clarkson and
A. A.
Coley\footnote{Email:~\texttt{clarkson@mathstat.dal.ca,~aac@mathstat.dal.ca.}}}
\address{Department of Mathematics and Statistics, Dalhousie University, Halifax,
        Nova Scotia, Canada, B3H~3J5.\\}

\date{\today}

\begin{document}
\maketitle

\begin{abstract}

Observations of the high degree of isotropy of the cosmic microwave background
are commonly believed to indicate that the Universe is `almost'
Friedmann-Lema\^\i tre-Robertson-Walker (at least since the time of last
scattering). Theoretical support for this belief comes from the so-called
Ehlers-Geren-Sachs Theorem. We show that a generalization of  this theorem
rules out any strong magnetic fields in the Universe. Our theoretical result
is model-independent and includes the case of inhomogeneous magnetic fields,
complementing previous results. We thus prove that cosmic microwave background
observations severely constrain all types of primordial and protogalactic
magnetic fields in the universe.

\end{abstract}

Magnetic fields seem to be abundant in the universe, and it is an open and
important question as to whether the origin of these fields are either
primordial (i.e. originating in the early universe and already present at the
onset of structure formation), or protogalactic (i.e. generated by battery
mechanisms during the initial stages of structure formation). One way to
distinguish between these two possibilities would be to detect or rule out the
presence of fields coherent on cosmological scales during recombination via
their imprint on the cosmic microwave background (CMB) radiation. Dynamically
significant magnetic fields present during recombination must be primordial.
Indeed, any {\it large-scale} primordial magnetic field with a strength
comparable to that inferred from the lowest measured intergalactic fields and
close to the observational upper limits via Faraday rotation measurements
\cite{x} may well be of cosmological origin. The detected magnetic field
strengths in high redshift galaxies \cite{Kron} and in damped Lyman alpha
clouds \cite{Wolfe} are consistent with the observed Faraday rotation
measurements. Primordial magnetic fields can be created in the early universe
through phase transitions or via inflation. However, even the invocation of
protogalactic dynamos to explain the magnitude of the field involves many
uncertain assumptions and still requires a small primordial (pregalactic) seed
field \cite{a}. Any seed magnetic field  is then amplified adiabatically during
gravitational collapse. Hence the possibility of a primordial field merits
serious consideration.

It is consequently of primary interest to determine the effects these magnetic
fields have on the CMB and, in particular, to consider the limits on any
large-scale primordial field from the CMB isotropy measurements. Indeed,  the
CMB measurements may lead to severe constraints on \emph{homogeneous} magnetic
fields in particular classes of cosmological models~\cite{BFS}. Here we
demonstrate quite generally that severe limits may be put on \emph{any} large
scale magnetic field~-- be it homogeneous, as considered before, or
inhomogeneous~-- from consideration of the high isotropy of the CMB. That is,
we show that any universe in which its observers measure an exactly isotropic
CMB cannot have a large-scale magnetic field. We show this by demonstrating
the magnetic field evolves in a manner which is irreconcilable with the
spacetime geometry required by the (exact) isotropy of the CMB. Contrary to
popular belief, this has not been done before.

We wish to take a form of the Copernican principle~-- which is the fundamental
assumption in most of cosmology~-- combined with our well established 
knowledge of the high isotropy of the CMB as our starting point. Thus we wish 
to consider the class of spacetimes which allow \emph{every} observer to see 
an isotropic CMB, and determine whether any of these are compatible with a 
large-scale magnetic field. Depending on additional assumptions that can be 
made, a surprising amount can be inferred from  this starting point; for 
example, if the matter in the spacetime is purely baryonic cold dark matter 
(`dust') with a contribution from radiation then the resulting spacetime must 
be Friedmann-Lema\^\i tre-Robertson-Walker (FLRW). This profound result is 
known as the Ehlers-Geren-Sachs theorem (EGS,~\cite{EGS,egsother}) and 
underpins the standard model of cosmology, not to mention many of the results 
of CMB physics. However, the Universe is not made up only of baryons but also 
dark matter, most of which could well be non-baryonic~\cite{carr}. In 
addition, in light of the supernovae Ia data~\cite{snia}, there could be a 
large contribution from a dynamic scalar field (e.g., 
quintessence~\cite{quint}), and there is no reason to believe that this would 
be homogeneous. Thus the EGS theorem is not the last word on the validity of 
the homogeneity of the standard model, and recent work has investigated 
certain generalizations of the EGS 
theorem~\cite{cla-bar99,fer-95,thesis,ccs,ccon} to inhomogeneous models. Most 
importantly for this work is the role a magnetic field will play in relation 
to these theorems.

In its most general form, the `isotropic radiation field' theorem identifies
spacetimes in which all observers see an `isotropic radiation field'. This may
be derived from the Einstein-Boltzmann equations for photons in a curved
spacetime~\cite{egsother,ellis}. It is then easy to show from the multipole
expansions of~\cite{ellis} that in a spacetime in which all observers on some
timelike congruence $u^a$ see an exactly isotropic radiation field, then this
velocity field has two important properties~\cite{EGS,cla-bar99}: First, the
expansion $\theta$ and acceleration $\udot_a$ (the motion of the observers
under non-gravitational forces) of these observers are related by
\be\label{irf}
\udot_a=\sdel_a Q,~~~\theta=3\dot Q,
\ee
where $Q$ is related to the energy density of  the radiation field. Here 
`$\sdel_a$' refers to the gradient in the instantaneous rest space of the 
observers (a generalization of the familiar `grad' of Euclidean 3-dimensional 
vector calculus), and `$~^\cdot~$' is a proper time derivative along the fluid 
flow. Second, the congruence must be shearfree, which means that the 
congruence cannot have any distortion (note that this condition is \emph{not} 
satisfied in the spacetimes of~\cite{BFS}). In many cases of physical interest 
it can then be proven that these observers have zero rotation~\cite{ccs} 
\footnote{For example, if part of the matter consists of a conserved comoving 
barotropic perfect fluid other than radiation, or for geodesic motion with any 
matter source, it follows from (1) that the expansion or the rotation must be 
zero. (For a conserved barotropic perfect fluid, we have 
$\dot{u}_a=\widetilde{\nabla}_a \phi$, and $p'\theta=\dot\phi$, where 
$\phi\equiv-\int dp/(\mu(p)+p)$, and $p'=dp/d\mu$; so, 
$\eta_{abc}\widetilde{\nabla}^b\widetilde{\nabla}^c(Q-\phi)= 
2(\frac{1}{3}-p')\theta\omega_a=0$. For geodesic motion, 
$\eta_{abc}\widetilde{\nabla}^b\widetilde{\nabla}^c 
Q=\frac{2}{3}\theta\omega_a=0$ [11].)},  a condition supported in part by 
isotropy of number-counts~\cite{rot}. For simplicity of exposition we shall 
examine only the irrotational case here. The general case of non-zero rotation 
is complicated, and will be considered elsewhere~\cite{ccs}.

Now, if we have a pure magnetic field~$B^a$, it will contribute an anisotropic
pressure, $\Pi_{ab}=-B_{\<a}B_{b\>}$\footnote{Angled brackets denote the
`projected, symmetric, trace-free' part of a tensor, where the projection is
into the observers' instantaneous rest space, and are also used to denote the
projected symmetric and trace-free part of time derivatives of spatial tensors
(i.e., the angled brackets are applied after the time
derivative)~\cite{maar-99,ellis}.} to the energy-momentum tensor of the
spacetime, as well as an energy density~$\mu_B=\:12B^2$, and an isotropic
pressure,~$p_B=\:13\mu_B$~\cite{mag,maa-tsa,ellis}.  The energy of the magnetic
field must be conserved, as required by one of the Einstein-Maxwell equations
(the induction equation), which is an evolution equation for the magnetic
field vector;
\be
\dot B_{\<a\>}=-\:23\theta B_a.
\ee
This implies that the
anisotropic pressure evolves as~\cite{maa-tsa}
\be\label{pi-dot}
\dot\Pi_{\<ab\>} =-\:43\theta\Pi_{ab}:
\ee
it is this evolution equation~-- which is a direct consequence of Maxwell's
equations~-- which is inconsistent with Einstein's field equations when an
isotropic radiation field is present, as we shall now show.

In the usual $1+3$ covariant approach, Einstein's field equations are broken
into a set of evolution and constraint equations~\cite{ellis}, which we
specialize to our particular case, to determine whether the magnetic field may
form part of the source. The evolution equation for the shear, in our case,
becomes an algebraic equation between the anisotropic pressures,~$\Pi_{ab}$,
and the electric part of the Weyl tensor,~$E_{ab}$, which governs tidal forces
and gravitational waves:
\be\label{E-Pi}
E_{ab}-\:12\Pi_{ab}=A_{ab},
\ee
where $A_{ab}=\sdel_{\<a}\udot_{b\>}+\udot_{\<a}\udot_{b\>}$ is a contribution
from acceleration terms. The electric Weyl tensor must satisfy the evolution
equation
\be\label{E-dot}
\dot E_{\<ab\>}+\theta E_{ab}=-\:12\bra{\dot\Pi_{\<ab\>}+\:13\theta\Pi_{ab}}.
\ee
(These equations may be found in e.g.,~\cite{maar-99,ellis} for a general
spacetime.) It may be shown that the acceleration terms~$A_{ab}$ evolve as
\be\label{A-dot}
\dot A_{\<ab\>}=-\:13\theta A_{ab}.
\ee
[This may be shown using~(\ref{irf}), the Ricci identities, and Eqns.~(5)
and~(14) from~\cite{maar-99}: i.e., the fact that there is no energy flux from
the magnetic field and the matter implies that the expansion is homogeneous
when the rotation is zero, which in turn implies that the acceleration vector
evolves parallel to the velocity vector.] Now, combining~(\ref{E-Pi}),
(\ref{E-dot}) and~(\ref{A-dot}) yields the following evolution equation for
the anisotropic pressure:
\be\label{pi-dot2}
\dot\Pi_{\<ab\>}=-\:23\theta\Pi_{ab}-\:23\theta A_{ab}.
\ee
This is the general evolution equation for the anisotropic pressure for an
irrotational spacetime without energy flux which allows an isotropic radiation
field. It is now easy to see that a magnetic field is inconsistent with this
evolution rate: a magnetic field in these spacetimes must
satisfy~(\ref{pi-dot}), which, upon substitution into~(\ref{pi-dot2}), and
using~(\ref{A-dot}) implies that
\be
\theta\Pi_{ab}=0:
\ee
i.e., the spacetime must be static (which is not relevant to cosmology), or the
magnetic field must  vanish (in which case the inhomogeneous models will be
those found in~\cite{cla-bar99}, and are not FLRW in general). Note that we
have made no assumptions on the other matter present, other than it have a
perfect fluid form; it needn't be barotropic, or homogeneous (both assumed
in~\cite{BFS}), or geodesic; it may also consist of many different components,
e.g., a perfect fluid and a scalar field (e.g., quintessence). No additional
assumption on the spacetime geometry has been made; this result is consequently
model-independent. \emph{Thus we see that, in quite general terms, a magnetic
field in the Universe is not compatible with an exactly isotropic CMB.}

However, it should be noted that we chose the magnetic field to be `pure' in
the frame of the radiation~-- i.e., `frozen into' the matter; if it were pure
in another frame (i.e., with respect to some other observers), the energy flux
(the Poynting vector arising from the electric field due to motion through the
magnetic field~\cite{tsa-cla00}) would not be zero and the result would not
follow directly from the above argument. Also, in principle, models which
contain other anisotropic matter stresses need to be considered separately
(e.g., if there were anisotropic pressures from some other source which
happened to cancel with those of the magnetic field, then this result would not
necessarily apply from the above calculation).


Of course, this result is a theoretical result based on perfect isotropy of the
CMB, and this magnetic-EGS theorem is not \emph{directly} applicable to the
real Universe, since the CMB temperature is not \emph{exactly} isotropic. The
original EGS theorem has been generalized   to the `almost EGS
theorem'~\cite{almost}, which states that if all fundamental observers measure
the CMB temperature to be almost isotropic during some time interval in an
expanding dust\footnote{Note that this assumption is crucial to the
theorem~\cite{bar-cla00,thesis}} universe, then the universe is described by an
almost FLRW model during this time interval. Therefore, there will be an
accompanying approximate result (an `almost' magnetic result) from which the
observed isotropy of the CMB will lead to severe constraints on the magnetic
field. This may be done by simply following the above proof, but now including
terms which are zero here but keeping them small: thus we may derive a specific
value for the maximum strength of any magnetic field from our knowledge of the
CMB anisotropies.\footnote{There will, however, be some additional assumptions
required  in such a calculation~\cite{almost}: for example, in an exact result
such as ours, a quantity which is zero (such as the shear) will have zero
derivatives; whereas, in the `almost' case where a quantity is small, it may
not necessarily have small derivatives; these terms must be kept small by
assumption in order to complete the calculation.} Indeed, Barrow
\etal~\cite{BFS} have shown that the constraints from the CMB isotropy
measurements may provide very strong limits on the strength of a homogeneous
component of a primordial magnetic field, and stronger than those imposed by
primordial nucleosythesis constraints. However, the results of~\cite{BFS} are
not generic and are model dependent (see~\cite{WCEH} for details\footnote{In
particular, we note that in general Bianchi VII$_h$ models, considered
in~\cite{BFS}, \emph{cannot} have a magnetic field as a source~\cite{HJ}.});
furthermore, they only apply to homogeneous magnetic fields. In particular, as
the models in~\cite{BFS} have shear, it is not clear as to whether the derived
constraints are really characteristic of the presence of the magnetic
field.\footnote{The authors consider Bianchi~VII$_h$ models, which are
homogeneous flat spacetimes with shear. Therefore the models will have an
anisotropic CMB, regardless of how one chooses the matter content, as a
consequence of the isotropic radiation field theorem, because it prohibits
spacetimes with shear. In their work, the matter was chosen to be a magnetic
field and hence was related to the shear in a particular fashion. It is not
clear to what extent the anisotropy of the CMB in these models constrain the
magnetic field through its particular coupling to the shear, and to what
extent it restricts magnetic fields in general.} Hence our theoretical result
(which is \emph{not} model dependent) complements their results very nicely
and provides the necessary theoretical background to such limits. Indeed, a
simple calculation shows that the constraints on magnetic fields from this
analysis are of a similar order of magnitude obtained in~\cite{BFS}, but an
accurate limit requires a separate detailed analysis~\cite{ccm}. We have thus
demonstrated that \emph{severe constraints may be placed on magnetic fields in
the Universe as a consequence of the extremely high isotropy of the CMB.} In
particular, our results also apply to inhomogeneous magnetic fields, and we
note that any inflationary scenario leading to significant large-scale
primordial magnetic fields would presumably result in magnetic inhomogeneities.

So far we have considered how constraints on \emph{large-scale} magnetic fields
may be derived from the CMB anisotropies. We have not, however, given
consideration to other types of magnetic field which may be present in the
Universe. Specifically, `tangled' magnetic fields~-- i.e., strong magnetic
fields which are inhomogeneous on small scales, but `average' to zero on large
scales~-- may also be constrained in a similar way. For example, consider a
bundle of light rays coming from the CMB surface to us of some angular size.
We have seen that a magnetic field will distort an isotropic radiation field,
so that if our bundle were to pass through a strong small-scale magnetic
field, it would produce temperature anisotropies in the CMB multipoles
corresponding to size of the tangled field. This may be used to place limits
on the strength and size of these inhomogeneous small-scale fields~\cite{ccm}.
With the coming launch of the next generation MAP and Planck satellites to
measure the high-order multipoles of the CMB spectrum, we may be able to place
stringent limits on these types of fields.

We have considered how constraints on magnetic fields may be derived from the
temperature anisotropies of the CMB. What, then, would be the implication of
discovering, by some other means, a large scale magnetic field in the 
Universe? It would follow immediately that the real Universe does not satisfy 
some or all of the assumptions used in this theorem. Most interestingly, it 
could be indicating that all observers in the Universe do not measure such 
high isotropy of the CMB; that is, that the Copernican principle does not 
hold. Thus we see an unexpected route for a physical test of the Copernican 
principle: looking for magnetic fields.

All of the discussion here has been confined to general relativity. It should
be noted that the situation might be different in alternative theories of
gravity. For example, in scalar-tensor theories of gravity and low-energy
effective theories derived from string theory, the high isotropy of the CMB
when combined with the Copernican principle implies that the Universe is
isotropic and homogeneous in the case of geodesic matter \cite{ccon}. However,
in string theory the electromagnetic field is coupled to the dilaton (unlike
in general relativity in which the electromagnetic field is governed by the
conformally invariant Einstein-Maxwell equations), and so the vacuum
fluctuations of the electromagnetic field can be significantly amplified by
accelerated growth of the dilaton in the pre-big-bang phase; in this case the
primordial magnetic field might be strong enough to seed galactic dynamo
effects and explain the origin of cosmic magnetic fields observed on galactic
and intergalactic scales \cite{GGV}. Therefore the CMB constraints will likely
be even more restrictive in alternative theories of gravity.

We would like to thank Richard Barrett, Roy Maartens and Christos Tsagas for
helpful discussions and comments. This work was supported, in part, by the
Natural Sciences and Engineering Research Council of Canada and The Royal
Society.


\begin{thebibliography}{99}

\bibitem{x}  K. T. Kim, P. C. Tribble, and P. P. Kronberg, Astrophys.~J.
{\it \ }{\bf 379}, 80 (1991); R. Perley and G. Taylor, Astron.~J. {\it \ }
{\bf 101}, 1623 (1991); P. P. Kronberg, Rep. Prog. Phys., {\it \ }{\bf 57},
325 (1994).

\bibitem{Kron}  P. P. Kronberg, J. J. Perry and E. L. Zukowski,
Astrophys.~J.{\it \ }%
{\bf 387}, 528 (1992).

\bibitem{Wolfe} A.~M. Wolfe, K. Lanzette, and A.~L. Oren, Astrophys.~J.
{\bf 388}, 17 (1992).

\bibitem{a}  R. Pudritz and J. Silk, Astrophys.~J.{\it \ }{\bf 342}, 650 (1989);
R. Kulsrud, R. Cen, J. P. Ostriker, and D. Ryu, 1997 {\it Astrophys. J.} {\bf 
480} 481.

\bibitem{BFS}  J. D. Barrow, P. G. Ferreira, and J. Silk,
Phys. Rev Lett. {\bf78}, 3610 (1997).

\bibitem{EGS} J. Ehlers, P.~Geren, and R.~K. Sachs, J. Math. Phys.
{\bf 9} 1344 (1964).

\bibitem{egsother}
G.~F.~R. Ellis, D.~R. Matravers, and R. Treciokas,
\newblock {Ann. Phys.} {\bf 150}, 455 (1983); {\bf 150}, 487 (1983)

\bibitem{carr} B.~J. Carr, {\it gr-qc/0008005} (2000).

\bibitem{snia} S.  {Perlmutter}, et~al.
\newblock { Astrophys. J.} {\bf 517}, 565 (1999);
 A.~G. Riess, et~al.
\newblock { Astrophys. J.} {\bf 116}, 1009 (1998);
M. Hamuy, et~al.
\newblock { Astrophys. J.} {\bf 112}, 2391 (1996).

\bibitem{quint}
I. Zlatev, L. Wang, and P. J. Steinhardt, Phys. Rev. Lett. {\bf 82}, 896
(1999).


\bibitem{cla-bar99}
C.~A. Clarkson and R.~K. Barrett,
\newblock { Class. Quantum Grav.} {\bf 16}, 3781 (1999).

\bibitem{fer-95}
J.~J. Ferrando, J.~A. Morales, and M. Portilla,
\newblock { Phys. Rev. D} {\bf 46(2)}, 578 (1992)

\bibitem{thesis} C.~A. Clarkson,  Ph.~D. Thesis, University of Glasgow (1999)
{\it astro-ph/0008089}.

\bibitem{ccs} C.~A. Clarkson, A.~A. Coley, E. O'Neill, and R.~A. Sussman. In preparation (2001).

\bibitem{ccon} C.~A. Clarkson, A.~A. Coley, and E. O'Neill. To appear in Phys. Rev. D (2000) 
\emph{gr-qc/0105026}.

\bibitem{ellis} G. F.~R.  Ellis and H. van Elst, in M. Lachieze-Rey (ed.),
{\em Theoretical and Observational Cosmology}, NATO Science Series, Kluwer
Academic Publishers~(1998) {\em gr-qc/9812046}

\bibitem{rot}
 A.~J. Fennelly, Astrophys. J. {\bf 207} 693 (1976);
 Y.~N. Obukhov,  {\it astro-ph/0008106} (2000).

\bibitem{maar-99}
R. Maartens, T. Gebbie,  and G.~F. R. Ellis,
\newblock {Phys. Rev. D} {\bf 59}, 083506 (1999).

\bibitem{mag} C.~G.  Tsagas and J.~D. Barrow, Class. Quantum Grav.
{\bf 15} 3523 (1998);
 {\bf 14} 2539 (1997).

\bibitem{maa-tsa} C.~G. Tsagas and R. Maartens,   Class. Quantum Grav.
{\bf 17} 2215 (2000).

\bibitem{tsa-cla00}
C.~G. Tsagas and C.~A. Clarkson. In preparation (2001);
 R. Maartens,  {\it astro-ph/0007352} (2000).

\bibitem{almost}
W. R. Stoeger, R. Maartens, and G. F.~R. Ellis,   Astrophys.~J.  {bf 443}, 1
(1995); R. Maartens,  G. F.~R. Ellis, and W.~R.  Stoeger, Phys. Rev. D {\bf
51}, 1525 (1995);
 {\bf 51}, 5942 (1995);  Astrophys.~J., {\bf 309}, L7 (1996).

\bibitem{bar-cla00} R.~K. Barrett and C.~A. Clarkson,
Class. Quantum Grav. {\bf 17} (2000) 5047–5078. {\it astro-ph/9911235}.

\bibitem{WCEH} J. Wainwright,  A.~A. Coley, G. F.~R. Ellis,  and M. Hancock,
Class. Quantum Grav. {\bf 15} 1331 (1998).

\bibitem{HJ} L.P. Hughston and K.C. Jacobs, Astrophys. J. {\bf 160} 147-52 (1970)


\bibitem{ccm}
C. A. Clarkson, A. A. Coley, R. Maartens, and C.~G. Tsagas. In preparation
(2001).

\bibitem{GGV}  M. Gasperini, M. Giovannini and G. Veneziano, Phys. Rev. Lett.
{\bf 75}, 3796 (1995).




\end{thebibliography}
\end{document}